\documentclass[twocolumn]{openjournal}

\usepackage{bm}
\usepackage{xspace}
\usepackage{xcolor}
\usepackage{amsmath}
\usepackage{xfrac}
\usepackage{graphicx}
\usepackage[hidelinks]{hyperref}
\hypersetup{breaklinks=true,colorlinks=true,linkcolor=blue,urlcolor=blue,citecolor=blue}
\usepackage[caption=false]{subfig}
\usepackage{booktabs}
\usepackage{orcidlink}

\usepackage{savesym}
\savesymbol{tablenum}
\usepackage{siunitx}
\restoresymbol{SIX}{tablenum}

\usepackage{tabularx}

\usepackage{lineno}

\usepackage{ragged2e}

\usepackage[capitalise]{cleveref}%must be added after amsmath

\creflabelformat{equation}{#2\textup{#1}#3}
\newcommand*{\rgw}{\ensuremath{r}}

\newcommand*{\BK}{BICEP/Keck}
\newcommand{\LCDM}{$\Lambda$CDM}

\newcommand*{\TT}{\ensuremath{T\!T}}
\newcommand*{\EE}{\ensuremath{E\!E}}
\newcommand*{\TE}{\ensuremath{T\!E}}

\newcommand*{\BB}{\ensuremath{B\!B}}
\newcommand*{\pp}{\ensuremath{\phi\phi}}

\newcommand*{\TTTEEE}{\ensuremath{T\!T/T\!E/E\!E}}

\newcommand*{\planck}{\textit{Planck}}

\newcommand{\RNum}[1]{\MakeUppercase{\romannumeral #1}}

\newcommand{\efold}{\ensuremath{N_{\star}}}

\newcommand{\ombh}{\ensuremath{\Omega_{\rm b} h^2}}
\newcommand{\omch}{\ensuremath{\Omega_{\rm c} h^2}}
\newcommand{\omegam}{\ensuremath{\Omega_{\rm m}}}
\newcommand{\As}{\ensuremath{A_{\mathrm{s}}}}
\newcommand{\logA}{\ensuremath{\log(10^{10}\,\As{})}}
\newcommand{\Hubble}{\ensuremath{H_0}}
\newcommand{\taureio}{\ensuremath{\tau_\mathrm{reio}}}
\newcommand{\ns}{\ensuremath{n_\mathrm{s}}}

\newcommand{\omm}{\omegam}

\newcommand{\rdrag}{\ensuremath{r_\mathrm{d}}}

\newcommand{\hrd}{\ensuremath{h\rdrag}}

\def\class{\texttt{CLASS}\xspace}

\def\cobaya{\texttt{Cobaya}\xspace}

\graphicspath{{./}{figures/}}

\shorttitle{Inflation at the End of 2025: Constraints on \rgw{} and \ns{} Using the Latest CMB and BAO Data}
\shortauthors{L. Balkenhol et al.}

\begin{document}

%\linenumbers
\journalinfo{The Open Journal of Astrophysics}

\title{Inflation at the End of 2025:\\ Constraints on \rgw{} and \ns{} Using the Latest CMB and BAO Data\vspace{-10ex}}

\author{L.~Balkenhol\,\orcidlink{0000-0001-6899-1873}$^{1,\ast}$}
\author{E.~Camphuis\,\orcidlink{0000-0003-3483-8461}$^{1}$}
\author{F.~Finelli\,\orcidlink{0000-0002-6694-3269}$^{2,3}$}
\author{K.~Benabed\,$^{1}$}
\author{F.~R.~Bouchet\,\orcidlink{0000-0002-8051-2924}$^{1}$}
\author{J.~Carron\,$^{4,5}$}
\author{S.~Galli\,$^{1}$}
\author{E.~Hivon\,\orcidlink{0000-0003-1880-2733}$^{1}$}
\author{A.~R.~Khalife\,\orcidlink{0000-0002-8388-4950}$^{1}$}
\author{L.~Knox\,$^{6}$}
\author{C.~L.~Reichardt\,\orcidlink{0000-0003-2226-9169}$^{7}$}
\author{A.~Vitrier\,\orcidlink{0009-0009-3168-092X}$^{1}$}
\author{W.~L.~K.~Wu\,\orcidlink{0000-0001-5411-6920}$^{8}$}
\email{$^\ast$lennart.balkenhol@iap.fr}

\affiliation{
$^1$Sorbonne Universit\'{e}, CNRS, UMR 7095, Institut d'Astrophysique de Paris, 98 bis bd Arago, 75014 Paris, France
}
\affiliation{
$^2$INAF/OAS Bologna, via Piero Gobetti 101, 40129 Bologna, Italy}
\affiliation{
$^3$INFN, Sezione di Bologna, viale C. Berti Pichat 6/2, 40127 Bologna, Italy}
\affiliation{$^4$Universit\'e de Gen\`eve, D\'epartement de Physique Th\'eorique et CAP, 24 Quai Ansermet, CH-1211 Gen\`eve 4, Switzerland}
\affiliation{$^5$Department of Physics \& Astronomy, University of Sussex, Brighton BN1 9QH, UK}
\affiliation{$^6$Department of Physics \& Astronomy, University of California, One Shields Avenue, Davis, CA 95616, USA}
\affiliation{$^7$School of Physics, University of Melbourne, Parkville, VIC 3010, Australia}
\affiliation{$^8$California Institute of Technology, 1200 East California Boulevard., Pasadena, CA, 91125, USA}

\begin{abstract}
Inflation elegantly provides initial conditions for the standard model of cosmology, while solving the horizon, flatness, and magnetic monopole problems.
Inflationary models make predictions for the tensor-to-scalar ratio \rgw{} and the spectral index \ns{} of initial density fluctuations.
In light of relevant data releases this year, we present constraints on these two parameters using the latest cosmic microwave background (CMB) and baryon acoustic oscillation data (BAO) available.
Using data from Planck, the South Pole Telescope, Atacama Cosmology Telescope, and BICEP/Keck experiments, we derive $\ns{}=0.9682\,\pm\,0.0032$ and a 95\% upper limit of $\rgw{}<0.034$.
This upper limit on \rgw{} is consistent with the official BICEP/Keck result given the numerical precision of the analyses and our choice to impose the self-consistency relation for single field slow-roll inflation on the tensor power spectrum;
the \rgw{} constraint is not impacted by the additional CMB data.
While adding DESI BAO data to the CMB data has a negligible impact on \rgw{}, the \ns{} constraint shifts upward to $0.9728\,\pm\,0.0029$, which favours monomial inflaton potentials with $\efold{}\!\sim 50$ over Starobinsky $R^2$ or Higgs inflation with $\efold{} = 51$ and $\efold{} = 55$, respectively.
This shift is caused by marginally significant differences between the CMB and DESI data that remain unexplained in the context of the standard model.
We show that a class of polynomial $\alpha$-attractor models can predict the CMB as well as the CMB+DESI \ns{} results. % with $\efold{}=47.1$ and $\efold{}=55.1$, respectively.
While future data will improve our sensitivity to \rgw{}, robust \ns{} constraints are just as crucial to differentiate between inflation models.
We make the data needed to reproduce the new CMB and BAO results and visualisation tools for \rgw{}-\ns{} figures to compare to any inflation model available \href{https://github.com/Lbalkenhol/r_ns_2025}{here}.
\end{abstract}

\maketitle

%%%%%%%%%%%%%%%%%%%%%%%%%%%%%%%%%%%%%%%%%%%%%%%%%%

%%%%%%%%%%%%%%%%% BODY OF PAPER %%%%%%%%%%%%%%%%%%

\section{Introduction}

Inflation, a period of near-exponential expansion in the early universe, offers an elegant explanation for the observed homogeneity and flatness of the universe and explains the absence of magnetic monopoles.
At the same time, if the inflaton is in its ground state, the simplest models of inflation predict small, Gaussian, adiabatic, close-to-scale-invariant perturbations, which serve as the initial conditions of the highly successful cosmological constant ($\Lambda$) cold dark matter (CDM) model of cosmology \citep{liddlecosmo, moderncosmo, weinbergcosmo, inflpostplanck13, abazajian16, baumanncosmo}.
Observations of the cosmic microwave background give strong evidence in support of these initial conditions; the CMB is remarkably Gaussian\footnote{with well-measured deviations due to the gravitational lensing by large-scale structure imprinting after recombination \citep{lewis06, planck18-8, qu24, ge24}} and measurements of its power spectrum constrain the spectral index of initial scalar perturbations, \ns{}, to better than percent precision, ruling out scale-invariance in favour of a red spectrum at $>10\,\sigma$ assuming the standard \LCDM{} model \citep{bennett13, planck18-1, planck18-6, planck18-10, louis25, camphuis25}.
However, inflation also predicts a background of gravitational waves, which has not been detected yet;
the most promising way of detecting these gravitational waves is through their signature in the $B$-mode polarisation of the CMB, as parametrised by the tensor-to-scalar ratio \rgw{} \citep{kamionkowski15, bicep2keck21b}.
%Still, direct observational evidence for inflation is missing.
%The most promising way of confirming this prediction is... finding such evidence is the search for the signature of gravitational waves produced by inflation in the $B$-mode polarisation of the CMB, as parametrised by the tensor-to-scalar ratio \rgw{} \citep{kamionkowski15, bicep2keck21b}.
In the absence of a detection, one can nevertheless discriminate between inflation models based on the upper limit on \rgw{} and precise determination of \ns{} that the available data allow.
%As it stands, different classes of inflationary models are compatible with the available data and succeed in producing the aforementioned conditions.

This year, three leading cosmological experiments, the South Pole Telescope (SPT) \citep{carlstrom11, camphuis25, quan26}, the Atacama Cosmology Telescope (ACT) \citep{fowler07, koopman16, naess25, louis25, calabrese25}, and the Dark Energy Spectroscopic Instrument (DESI) \citep{desi13, desi16b, desi22, desi25}, released data sets that, alone or in combination with one another, are sensitive to \ns{}.
In this work, we combine these data sets with \planck{} data \citep{planck18-1, carron22} and the leading CMB $B$-mode data of \BK{} \citep{bicep2keck21b} to present the state of \rgw{}-\ns{} constraints at the end of 2025.
This work is not intended to be a comprehensive review of constraints in the inflationary model space, but instead has one singular goal: to present the latest constraints on \rgw{} and \ns{} as a reference for the community.
Accordingly, we make the data products associated with these results publicly available.
Additionally, we release visualisation scripts and an online application that allow for the creation of custom \rgw{}-\ns{} figures with data constraints and the prediction of any inflationary model of choice, with the intention that these resources remain useful as new data arrive and models are developed.\footnote{\url{https://github.com/Lbalkenhol/r_ns_2025},\\ \url{https://r-ns-plot.streamlit.app/}}

This work is structured as follows.
In \S\ref{sec:data_method} we specify the data used, the model space considered, and our methodology to derive parameter constraints.
In \S\ref{sec:results} we present our results, before concluding in \S\ref{sec:conclusion}.

%%%%%%%%%%%%%%%%%%%%%%%%%%%%%%%%%%%%%%%%%%%%%%%%%%%%%%%%%%%%%%
\section{Data Sets and Methodology}
\label{sec:data_method}

% Model
We present constraints on the \LCDM{} model of cosmology, adding a varying tensor-to-scalar ratio $\rgw$ (at a pivot scale of $0.05\, \mathrm{Mpc}^{-1}$).
We set the tensor power spectrum tilt to obey the self-consistency condition for single field slow-roll inflation.
%\footnote{We set the tensor power spectrum tilt and its running at the pivot scale to obey the self-consistency condition for single field slow-roll inflation.}
We otherwise parametrise the model using the amplitude \logA{} and spectral index \ns{} of initial scalar perturbations (at the same pivot scale as \rgw{}), the physical density of baryons \ombh{} and dark matter \omch{}, the expansion rate today \Hubble{}, and the optical depth to reionisation \taureio{}.

% Data
To constrain this model, we use CMB data from \planck{}, SPT, ACT, and \BK{}.
Specifically, for the first three experiments, we use the `SPA' combination of primary CMB (\TTTEEE{}) and CMB lensing (\pp{}) measurements introduced by \citet[][see \S\RNum{7}A therein]{camphuis25} featuring SPT-3G D1, ACT DR6, as well as \planck{} PR3 and PR4 data \citep{planck18-1, carron22, qu24, madhavacheril24, ge24, balkenhol24, camphuis25, naess25, louis25, calabrese25, quan26}.
As introduced by \citet{louis25}, and matching the definition of `SPA' in \citet{camphuis25}, we restrict the \planck{} data to $\ell < 1000$ in \TT{} and $\ell < 600$ in \TE{}/\EE{} to avoid double counting information.
Note that we use the compressed CMB-only likelihoods, where available.
We use the latest \BK{} $B$-mode (\BB{}) likelihood \citep{bicep2keck21b}.
We also consider BAO measurements from the second data release (DR2) of the DESI collaboration \citep{desi25}.
We consider data sets that agree across the common constrained parameters with an equivalent one-dimensional Gaussian significance of $3\,\sigma$ or less as consistent and allow ourselves to combine them;
this has been demonstrated for the different CMB data sets and for the SPA combination w.r.t the BAO data \citep{louis25, camphuis25}.

% Sampling
We perform Markov Chain Monte Carlo (MCMC) analyses using the \cobaya{} sampler \citep{lewis02b, neal05, bobyqa, lewis13eff, cartis18, cartis18b, torrado21} and consider runs with a Gelman-Rubin statistic of $R-1<0.02$ as sufficiently converged.
To obtain model predictions, we use the Boltzmann solver \class{} \citep{blas11}.
As the CMB data set combination above does not include large-scale $E$-mode data, we impose a \planck{}-based prior of $\mathcal{N}\sim(0.051, 0.006^2)$ on the optical depth to reionisation \taureio{} \citep{planck20-57}, matching the definition of `SPA' in \citet{{camphuis25}}.
We otherwise impose uniform priors on cosmological parameters.
For \rgw{} we report 95\% upper limits, whereas for \ns{} we report mean values, 68\% confidence intervals, and best-fit values in parentheses.
Figures are generated with the help of \texttt{GetDist} \citep{getdist25}.

%\begin{table}
%\renewcommand{\arraystretch}{1.5}
%\centering
%\begin{tabular}{l c c}
%Data set & $\rgw{}$ & $\ns{}$ \\
%\hline
%SPA+BK & $0.034$ & $0.9681\,\pm\,0.0032\,(0.9681)$ \\
%SPA+BK+DESI & $0.035$ & $0.9727\,\pm\,0.0030\,(0.9728)$ \\
%\end{tabular}
%\caption{95\% upper limits on \rgw{} along mean values and 68\% confidence intervals for \ns{} derived from different combinations of CMB and BAO data (see \S\ref{sec:data_method} for details on the data sets).
%The best-fit \ns{} values are given in parentheses.}
%\label{tab:rns}
%\end{table}

\begin{figure*}[ht!]
    \includegraphics[width=2.0\columnwidth]{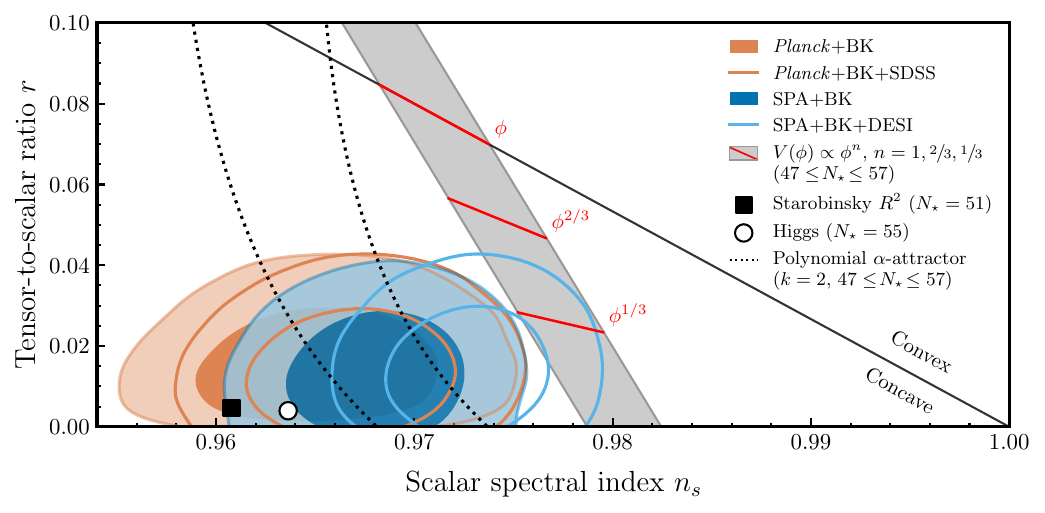}
    \caption{Constraints on the tensor-to-scalar ratio \rgw{} and the scalar spectral index \ns{} from different cosmological data sets alongside theoretical predictions.
    The orange contours are derived from \planck{} and \BK{} data, whereas the blue contours additionally use SPT and ACT data.\footnote{Note that for the orange contours, \planck{} low-E data and PR3 lensing data are used, whereas for the blue contours, we use \planck{} PR4 lensing data and a prior on the optical depth to reionisation is applied in lieu of the \planck{} low-E data. Both orange contours use \planck{} multi-frequency likelihoods. The open orange contours use the public data products of \citet{bicep2keck21b}.}
    Filled contours use only CMB data, whereas line contours add BAO data (SDSS for orange contours, DESI for blue contours).
    We show the prediction for monomial inflation models with different exponents in red; the grey shaded region corresponds to a duration of inflation of \efold{}$=47$-$57$ e-folds for monomial potentials.
    The predictions of Higgs inflation for $\efold{}=55$ and Starobinsky $R^2$ inflation for $\efold{}=51$ are shown as a white circle and a black square, respectively.
    The grey dotted lines correspond to the predictions of a polynomial $\alpha$-attractor model with potential $V(\varphi) = V_0 |\varphi|^2/(\mu^2 + |\varphi|^2)$ for $\efold=47$-$57$ (left to right dotted line).
    %and an example of the polynomial $\alpha$-attractor class with potential $V(\varphi) = V_0 |\varphi|^2/(\mu^2 + |\varphi|^2)$ for $\efold{}=55,\, \mu\approx0.87$ are shown as a white circle, a black square and a white diamond, respectively.
    %The predictions of Higgs inflation for $\efold{}=55$ and Starobinsky $R^2$ inflation for $\efold{}=51$ are shown as a white circle and a black square, respectively.
    We divide the \rgw{}-\ns{} plane into regions of concave and convex inflaton potentials.
    The code to reproduce this figure with SPA+BK(+DESI) contours and all shown theoretical predictions is provided at \url{https://github.com/Lbalkenhol/r_ns_2025}.
    %\citep{Bezrukov12, bicep2keck21b, lb23}.
    }
    \label{fig:rns_data}
\end{figure*}

%%%%%%%%%%%%%%%%%%%%%%%%%%%%%%%%%%%%%%%%%%%%%%%%%%%%%%%%%%%%%%
\section{Results}
\label{sec:results}

We now report constraints on \rgw{} and \ns{} from the data sets introduced in the previous section.
We show marginalized posterior distributions in the $\rgw{}$-$\ns{}$ plane in Figure \ref{fig:rns_data} alongside different theoretical predictions.
%Table \ref{tab:rns} summarizes the \rgw{} and \ns{} constraints.
There exists a vast and changing landscape of inflation models \citep[for recent reviews see][]{martin13, kallosh25} and an exhaustive comparison of the data constraints with the spectrum of theoretical predictions available is beyond the scope of this work;
we instead choose to compare to a set of reference models, namely monomial inflaton potentials with $\efold{}=47$-$57$ e-folds, Starobinsky $R^2$ inflation with $\efold{}=51$ \citep{starobinsky80, mukhanov81, staro83}, Higgs inflation with $\efold{}=55$ \citep{higgs08, Bezrukov12},\footnote{We choose to fix the number of e-folds for Starobinsky and Higgs inflation for simplicity;
for these models a range of \efold{} values is allowed reflecting the uncertainty on reheating.} and polynomial $\alpha$-attractor models with potential $V(\varphi) = V_0 |\varphi|^k/(\mu^k + |\varphi|^k)$ with $k=2$ and $47\leq \efold{} \leq 57$ \citep{kallosh22}.\footnote{For the chosen polynomial $\alpha$-attractors \ns{} depends on $k$ and \efold{}, while \rgw{} additionally also depends on $\mu$. As only upper limits have been reported for \rgw{}, we focus our discussion on \ns{} for this model, which is given by $\ns=1-2(k+1)/\left[\efold{}(k+2)\right]$ in the limit of small $\mu$ and large \efold{}. For details on the different classes of $\alpha$-attractor models see \citet{kallosh25s}.}
%and a polynomial $\alpha$-attractor model with potential $V(\varphi) = V_0 |\varphi|^k/(\mu^k + |\varphi|^k)$ with $k=2$, $\efold{}=55$, and $r=3\times10^{-3}$ \citep{kallosh22}.\footnote{For polynomial $\alpha$-attractors \ns{} depends on $k$ and \efold{}, whereas \rgw{} additionally also depends on $\mu$. As only upper limits on \rgw{} have been reported, we focus our discussion on \ns{} constraints for this model and choose to set \rgw{} to coincide with the fiducial point of the forecast discussed later in this section, $\rgw{}=3\times10^{-3}$, which corresponds to $\mu\approx0.87$.}
Here we restrict ourselves to $(\ns{}, \rgw{})$ predictions to lowest order in slow-roll parameters, which is consistent with the Higgs and Starobinsky models differing only by the reheating temperature, i.e. by \efold{};
we caution the reader about the contribution of next-to-leading terms in slow-roll parameters might be non-negligible in our discussion.
The data products and code made publicly available with this work can be used to compare the new data constraints to arbitrary predictions for \rgw{} and \ns{}.

We first consider constraints from the SPA+BK combination.
We report:
\begin{align}
\rgw{} &< 0.034,\notag\\ 
\ns{} &= 0.9682\,\pm\,0.0032\,(0.9681).
\end{align}
These constraints are consistent with the prediction of Starobinsky $R^2$ and Higgs inflation at $2.0\,\sigma$ and $1.3\,\sigma$, respectively.
They are compatible with monomial potentials with $\efold{} = 47$ at $2.0\,\sigma$ (for $V(\phi)\propto\phi^n$ with $n=0.33$) and differ from the closest convex potential at $4.8\,\sigma$.
The \ns{} central value matches the prediction of the chosen polynomial $\alpha$-attractor model for $\efold{}=47.1$ and remains compatible with the prediction for $\efold{}=57$ at $1.7\,\sigma$.
%is compatible with the prediction of the chosen polynomial $\alpha$-attractor model at $<0.1\,\sigma$ and $1.7\,\sigma$ for $\efold{}=47$ and $\efold{}=57$, respectively.
%Note that these indicative exclusion ranges refer to the particular $\efold{}$ values adopted to represent these inflationary models and do not fully take into account reheating uncertainties.
Note that these indicative exclusion ranges refer to the particular $\efold{}$ values adopted to represent these inflationary models and do not fully take into account reheating uncertainties as well as next-to-leading terms in slow-roll parameters in the $(\ns{}, \rgw{})$ predictions.

The constraint on \rgw{} is dominated by the \BK{} data; the minor change in the upper limit above compared to the $\rgw{}<0.035$ result of \citet{bicep2keck21b} (see Figure 5 therein for the combination of \BK{}, \planck{}, and SDSS BAO data, as well as \citet{paoletti22}) is compatible with the numerical precision of the MCMC analyses given the effective number of \rgw{} samples.
Additionally, we note that \citet{bicep2keck21b} fix the tensor power spectrum tilt to zero, whereas in our analysis it is set using the self-consistency relation for single field slow-roll inflation.
This tends to slightly increase the expected B-mode power and can hence lower the derived upper limit \citep[see e.g.][]{seljak96bb, moderncosmo}.
The additional CMB data compared to \citet{bicep2keck21b} is expected to have no measurable effect on \rgw{}.
Note that in comparison to \citet{tristram22}, who report $\rgw{}<0.032$, we use a different combination of \planck{} data and in particular do not include \planck{} \BB{} spectra.

The \ns{} constraint on the other hand is informed by the \planck{}, ACT, and SPT data, with the data sets' sensitivity descending in this order due to their weighting across angular scales.
The individual CMB experiments' constraints on \ns{} (and \LCDM{} parameters more broadly) are consistent with one another, while being largely independent:\footnote{The SPT data set is derived from observations of 4\% of the sky and its overlap with the footprints of the other experiments is minimal \citep{camphuis25, qu25}. There is some mild correlation between the ACT and \planck{} data, though note that in the SPA combination the scales covered by both experiments are removed from the \planck{} data \citep{louis25}.} $\ns{}=0.9657\,\pm\,0.0040$ for \planck{}, $\ns{}=0.9682\,\pm\,0.0069$ for ACT, and $\ns{}=0.951\,\pm\,0.011$ for SPT.
Adding the SPT to \planck{} data yields $\ns{}=0.9636\,\pm\,0.0035$, whereas a joint analysis of SPT and ACT data gives $\ns{}=0.9671\,\pm\,0.0058$ \citep[all results taken from][]{camphuis25}.
Note that \citet{louis25} report $\ns{}=0.9709\,\pm\,0.0038$ for the P-ACT combination of \planck{} and ACT primary CMB data defined therein;
this is higher than the individual \planck{} and ACT results as differences in the baryon density between the ACT and cut \planck{} data are accommodated in the joint constraints by raising \ns{} (see the lower left panel of Figure 37 in \citet{louis25}).\footnote{The referenced analyses do not vary \rgw{} and do not include \BK{} data, though given the small correlation between these two parameters the impact is negligible. See also \citet{mcdonough25} and Table I of \citet{ellis25} for a compilation of \ns{} constraints as well as \citet{jense25planck} for details on the choice of \planck{} likelihood.}
CMB results show a high degree of robustness across \LCDM{} parameters when considering only temperature or polarisation data and when comparing constraints from different angular scales \citep{planck18-5, louis25, camphuis25}.\footnote{Note that the modelling of the Sunyaev–Zeldovich power spectrum can mildly impact the ACT constraint on \ns{} \citep{beringue25}. For more information on the stability of recent CMB results with respect to foreground modelling see \citet{tristram25}.}

\begin{figure*}[ht!]
    \includegraphics[width=2.0\columnwidth]{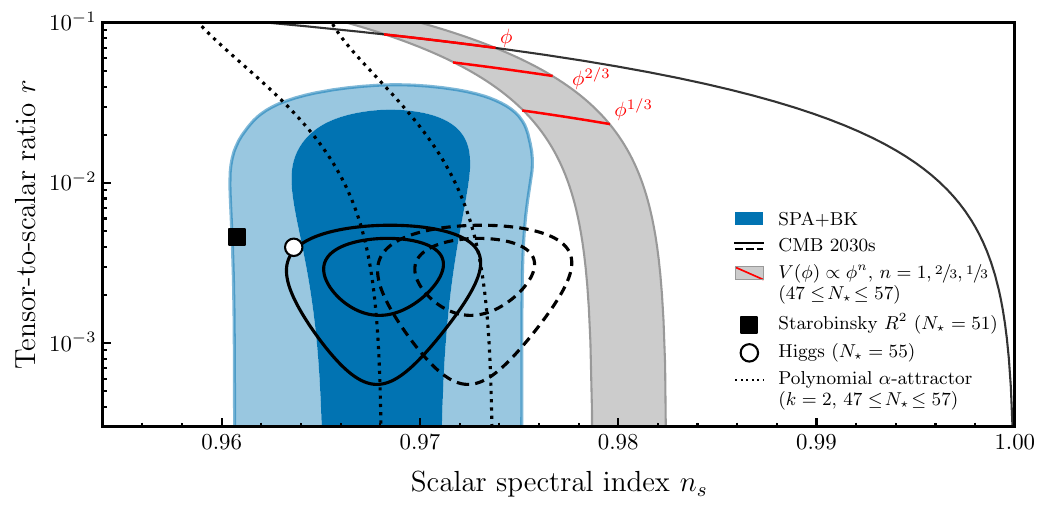}
    \caption{Constraints from SPA+BK on \rgw{} and \ns{} (blue contours, same as in Figure \ref{fig:rns_data}).
    Alongside the data constraints we show two forecasts for CMB constraints that may be achieved in the next decade ($10^3 \rgw{}=3\,\pm\,1,\,\sigma(\ns)=2\times 10^{-3}$) at different values of \ns{}: one at the central value of SPA+BK (black solid contours) and one at the central value of SPA+BK+DESI (black dashed contours).
    The theoretical predictions shown are the same as in Figure \ref{fig:rns_data}.
    }
    \label{fig:rns_forecast}
\end{figure*}

The SPA constraint on \ns{} is correlated at the $31\%$ level with \taureio{} and is as such sensitive to the prior chosen for the latter.
Using a symmetrised \texttt{Sroll2} prior of $\mathcal{N}\sim(0.0566, 0.0058^2)$ \citep{louis25, sroll2} instead of our baseline choice, we find $\ns{} = 0.9691\,\pm\,0.0033$;
the \ns{} central value shifts upward by approximately $0.3\,\sigma$, where $\sigma$ is the width of the baseline \ns{} posterior.
Completely removing the prior on \taureio{} leads to a constraint of $\taureio{} = 0.076\,\pm\,0.014$ (in line with the findings of \citet{camphuis25}).
As expected from the correlation of \taureio{} and \ns{}, in response the \ns{} posterior shifts towards larger values and widens to $\ns{} = 0.9722\,\pm\,0.0041$.
While raising the optical depth to reionisation to around $0.09$ (and consequently \ns{} to $\sim 0.9746$) is interesting in the context of neutrino mass constraints, such a high value is at odds with the \planck{} low-E data and there exists no known systematic effect in this measurement large enough to explain a shift of this size \citep{planck18-6, craig24, green25, sailer25, jhaveri25, camphuis25}.\footnote{See also \S IV of \citet{mcdonough25} for a discussion of \ns{} constraints from ACT data with a \taureio{} prior centred on $0.11$.}
Moreover, the combination of data from the CLASS telescope and the WMAP satellite does not support such a high value of \taureio{} either ($\taureio{} < 0.07$ at $95\%$ confidence) \citep{li25}.

We now add DESI to the CMB data.
For the SPA+BK+DESI combination we report:
\begin{align}
\rgw{} &< 0.035,\notag\\ 
\ns{} &= 0.9728\,\pm\,0.0029\,(0.9728).
\end{align}
As has been noted in the literature, the addition of DESI data shifts the central value of the \ns{} constraint upward \citep[see also references within these works]{louis25, camphuis25, ferreira25, mcdonough25, iacconi25}.
The central value of the \ns{} constraint coincides with the value predicted by the chosen polynomial $\alpha$-attractor model for $\efold{}=55.1$ and therefore this class of models is very consistent with the addition of DESI DR2 BAO data.
%The central value of the \ns{} constraint coincides with the value predicted by the chosen polynomial $\alpha$-attractor model for $\efold{}=55.1$; the result agrees with the predictions for $\efold{}=47$ and $\efold{}=57$ at $1.6\,\sigma$ and $0.3\,\sigma$, respectively.
%For the chosen polynomial $\alpha$-attractor model the \ns{} central value translates to a constraint on the number of e-folds of $\efold{}=55.7\,\pm\,6.2$.
%The \ns{} constraint coincides with the value predicted by the chosen polynomial $\alpha$-attractor model at $<0.1\,\sigma$.
%;taking also \rgw{} into account, this corresponds to a $0.8\,\sigma$ difference, though we stress that the model prediction for \rgw{} changes depending on the chosen value for $\mu$.
Taking also \rgw{} into account, the data constraints are compatible with a range of monomial inflaton potentials at $<2\,\sigma$ (including for $\efold{}>47$) as shown in Figure \ref{fig:rns_data}.
%While Higgs inflation remains compatible at $2.9\,\sigma$ with the data, the prediction of Starobinsky $R^2$ inflation differs from the data at the $3.9\,\sigma$ level.\footnote{We stress that this is for the chosen fiducial values of \efold{}. For both models, Starobinsky $R^2$ and Higgs inflation, the predicted \ns{} value can change depending on the details of reheating. Also note that for certain inflation models too few e-folds can result in reheating temperatures that are inconsistent with Big Bang Nucleosynthesis \citep{martin13, Zharov25}.}
Whereas the prediction of Higgs inflation remains compatible at $2.9\,\sigma$ with the data, the prediction of Starobinsky $R^2$ inflation differs from the data at the $3.9\,\sigma$ level.\footnote{We stress that this is for the theoretical predictions to lowest order in slow-roll parameters and for the chosen fiducial values of \efold{}. For both models, Starobinsky $R^2$ and Higgs inflation, the predicted \ns{} value can change depending on the details of reheating.
Also note that next-to-leading order terms in slow-roll parameters increase 
the value of  \ns{} for the Starobinsky model \cite{martin16}, partially reconciling the model with the addition of DESI, and that for certain inflation models too few e-folds can result in reheating temperatures that are inconsistent with Big Bang Nucleosynthesis \citep{martin13, Zharov25}.}
The closest convex monomial potential is at $3.9\,\sigma$ from the SPA+BK+DESI constraint.
%The data constraints are broadly consistent with the predictions of models belonging to the class of polynomial alpha attractors \citep{kallosh19, ferreira25}.

The shift in the \ns{} central value seen above occurs when DESI data is added, because the CMB \ns{} constraint is correlated with parameters well-determined by the BAO data: at $-57\%$ with the matter density \omm{} and $57\%$ with the product of the expansion rate and the size of the sound horizon at the end of the baryon drag epoch \hrd{}.\footnote{In \LCDM{}, \omm{} and \hrd{} represent a lossless compression of BAO data.}
Since the SPA and DESI results are compatible at the $2.8\,\sigma$ level in the \omm{}-\hrd{} plane \citep[][which we also confirm here]{camphuis25}, the correlation of these parameters with \ns{} provides a lever arm for the DESI data to change the inferred \ns{} value.
Note that due to the different weighting of CMB data sets across angular scales and spectra this effect does not always occur.
Specifically, for the SPT-3G D1 data set the anticorrelation between \ns{} and \omm{} is $5\,\%$ and hence the SPT \ns{} constraint of $\ns{}=0.951\,\pm\,0.011$ is negligibly changed by the addition of DESI data to $\ns{}=0.949\,\pm\,0.012$ \citep{camphuis25}, even though the separation of the constraints of the two data sets in the \omm{}-\hrd{} plane corresponds to $2.5\,\sigma$.
Note that when using the \texttt{Sroll2} \taureio{} prior instead of our baseline choice the SPA and DESI results are compatible at the $2.5\,\sigma$ level in the \omm{}-\hrd{} plane and hence joint constraints remain susceptible to this effect.
While the differences between CMB and DESI data may be statistically normal, they could also point to systematic errors or a failure of the standard model; we should hence interpret the conclusions regarding \ns{} with caution \citep{ferreira25}.

Future CMB data will yield improved constraints on \rgw{} and \ns{}.
The South Pole Observatory (the combination of the \BK{} and SPT experiments), the Simons Observatory, and the LiteBIRD mission all aim to achieve near $\sigma(\rgw{})=10^{-3}$ in the next decade \citep[][\BK{} Collaboration in prep.]{lb23, simonsobservatorycollab19, sofc2025, so_bb_new}.
At the same time, adding data from these missions to \planck{} data will also allow to reduce the uncertainty on the scalar spectral index to $\sigma(\ns{})\approx2\times 10^{-3}$ through improved measurements of the polarisation and lensing power spectra \citep{simonsobservatorycollab19, sofc2025, lb23, prabhu24, vitrier25}.
Together, these improvements lead to a reduction in the allowed \rgw{}-\ns{} parameter volume (using the figure of merit) by a factor of $16$ compared to the SPA+BK constraints above.

In Figure \ref{fig:rns_forecast} we show how the anticipated constraints of $\sigma(\rgw{})=10^{-3},\,\sigma(\ns)=2\times 10^{-3}$ compare to different theoretical predictions.
We centre the constraints on $3\times10^{-3}$ for \rgw{} and for \ns{} either on the central value of the SPA+BK or SPA+BK+DESI constraints.
While the improved sensitivity to \rgw{} is invaluable, robust \ns{} constraints are just as important to confidently explore inflationary models.
To illustrate this point, still to lowest order in slow-roll parameters and without taking into account reheating uncertainties, the compatibility of the anticipated constraints with the prediction for Higgs (Starobinsky) inflation changes from $2.0\,(3.7) \,\sigma$ to $4.4\, (6.0) \sigma$ depending on the adopted central value for \ns{}.
%To illustrate this point, the compatibility of the anticipated constraints with the prediction for Higgs inflation changes from $2.0\,\sigma$ to $4.4\,\sigma$ depending on the adopted central value for \ns{}.
%For the higher \ns{} central value Starobinsky inflation would be disfavoured at $6.0\,\sigma$; this would be a significant change to our understanding of inflation compared to the conclusions drawn following the \planck{} mission \citep{planck18-10, bicep2keck21b, ferreira25}.
Among the range of \ns{} values predicted by the chosen class of polynomial $\alpha$-attractors for $47\leq\efold{}\leq 57$, there are models compatible with both forecasts at $<1\,\sigma$.
%The \ns{} value predicted by the chosen polynomial $\alpha$-attractor model is compatible with the forecasts at $<0.1\,\sigma$ and $2.3\,\sigma$ for the high and low \ns{} central values, respectively.
The forecasts above do not take future BAO data into account.
%This forecast does not account the fact that future BAO data may allow for a further tightening of the \ns{} constraint, though it is unclear whether they will be consistent with the CMB data or increase the differences between the two probes further.

\section{Conclusions}\label{sec:conclusion}

We have presented the state of \rgw{} and \ns{} constraints at the end of 2025, using the latest CMB and BAO data available.
While the CMB data alone favour a region of the model space compatible with Higgs and Starobinsky $R^2$ inflation at $\leq 2.0\,\sigma$ (assuming $\efold{} = 55$ and $\efold{} = 51$, respectively), the addition of DESI data shifts the inferred \ns{} value high, in favour of monomial inflaton potentials (with an exponent $\sim\sfrac{1}{3})$.
This shift in \ns{} when BAO data are added is caused by marginally statistically significant differences between CMB and DESI data, which remain unexplained in the context of the standard model.
Classes of polynomial $\alpha$-attractor models are able to predict \ns{} values in agreement with the CMB and CMB+BAO results depending on the number of e-folds.
Future CMB data will improve our sensitivity to \rgw{} and \ns{}.
Further data from DESI as well as large-scale structure data from the Euclid satellite \citep{mellier24} may either increase the differences between CMB and BAO data to statistical significance, or resolve them and thus allow for more precise and robust analyses of inflationary models.

\begin{acknowledgments}

This work is dedicated to our colleague and friend Karim Benabed.
His brilliance and generosity remain forever inspirational.\\
L.B. is grateful to Tom Crawford for helpful discussions and encouragement and thanks Ahmed Soliman for feedback on the manuscript.
This project has received funding from the European Research Council (ERC) under the European Union’s Horizon 2020 research and innovation programme (grant agreement No 101001897).
W.L.K.W. acknowledges support from an Early Career Research Award of the Department of Energy.
C.L.R.  acknowledges support from the Australian Research Council’s Discovery Project scheme (No. DP210102386).
L.K. is supported in part by the Michael and Ester Vaida Endowed Chair in Cosmology and Astrophysics.
This work has received funding from the Centre National d’Etudes Spatiales and has made use of the Infinity Cluster hosted by the Institut d’Astrophysique de Paris. 
The South Pole Telescope program is supported by the National Science Foundation (NSF) through awards OPP-1852617 and OPP-2332483.
Partial support is also provided by the Kavli Institute of Cosmological Physics at the University of Chicago.

\end{acknowledgments}

%%%%%%%%%%%%%%%%%%%% REFERENCES %%%%%%%%%%%%%%%%%%

\bibliographystyle{aa}
\typeout{}
\bibliography{spt}

@ARTICLE{martin16,
       author = {{Martin}, J{\'e}r{\^o}me and {Ringeval}, Christophe and {Vennin}, Vincent},
        title = "{Shortcomings of new parametrizations of inflation}",
      journal = {\prd},
         year = 2016,
        month = dec,
       volume = {94},
       number = {12},
          eid = {123521},
        pages = {123521},
          doi = {10.1103/PhysRevD.94.123521},
archivePrefix = {arXiv},
       eprint = {1609.04739},
 primaryClass = {astro-ph.CO},
       adsurl = {https://ui.adsabs.harvard.edu/abs/2016PhRvD..94l3521M}
}

@ARTICLE{quan26,
       author = {{Quan}, W. and {Camphuis}, E. and {Daley}, C. and {Huang}, N. and {Omori}, Y. and {Guidi}, F. and {Anderes}, E. and {Anderson}, A.~J. and {Ansarinejad}, B. and {Archipley}, M. and {Balkenhol}, L. and {Barron}, D.~R. and {Benabed}, K. and {Bender}, A.~N. and {Benson}, B.~A. and {Bianchini}, F. and {Bleem}, L.~E. and {Bocquet}, S. and {Bouchet}, F.~R. and {Campitiello}, M.~G. and {Carlstrom}, J.~E. and {Carron}, J. and {Chang}, C.~L. and {Chichura}, P.~M. and {Chokshi}, A. and {Chou}, T.-L. and {Coerver}, A. and {Crawford}, T.~M. and {de Haan}, T. and {Dibert}, K.~R. and {Dobbs}, M.~A. and {Doohan}, M. and {Dutcher}, D. and {Feng}, C. and {Ferguson}, K.~R. and {Ferree}, N.~C. and {Fichman}, K. and {Foster}, A. and {Galli}, S. and {Gambrel}, A.~E. and {Gao}, A.~K. and {Ge}, F. and {Guns}, S. and {Halverson}, N.~W. and {Hivon}, E. and {Holder}, G.~P. and {Holzapfel}, W.~L. and {Hood}, J.~C. and {Hryciuk}, A. and {Jhaveri}, T. and {K{\'e}ruzor{\'e}}, F. and {Khalife}, A.~R. and {Knox}, L. and {Kornoelje}, K. and {Kuo}, C.-L. and {Levy}, K. and {Li}, Y. and {Lowitz}, A.~E. and {Lu}, C. and {Lynch}, G.~P. and {Maccarone}, T.~J. and {Maniyar}, A.~S. and {Martsen}, E.~S. and {Menanteau}, F. and {Millea}, M. and {Montgomery}, J. and {Nakato}, Y. and {Natoli}, T. and {Ouellette}, A. and {Pan}, Z. and {Paschos}, P. and {Phadke}, K.~A. and {Pollak}, A.~W. and {Prabhu}, K. and {Raghunathan}, S. and {Rahimi}, M. and {Rahlin}, A. and {Reichardt}, C.~L. and {Rouble}, M. and {Ruhl}, J.~E. and {Silva Oliveira}, A.~C. and {Simpson}, A. and {Sobrin}, J.~A. and {Stark}, A.~A. and {Stephen}, J. and {Tandoi}, C. and {Trendafilova}, C. and {Vieira}, J.~D. and {Vieregg}, A.~G. and {Vitrier}, A. and {Wan}, Y. and {Whitehorn}, N. and {Wu}, W.~L.~K. and {Young}, M.~R. and {Zebrowski}, J.~A.},
        title = "{SPT-3G D1: Maps of the millimeter-wave sky from 2019 and 2020 observations of the SPT-3G Main field}",
      journal = {arXiv e-prints},
         year = 2026,
        month = mar,
          eid = {arXiv:2603.20163},
        pages = {arXiv:2603.20163},
          doi = {10.48550/arXiv.2603.20163},
archivePrefix = {arXiv},
       eprint = {2603.20163},
 primaryClass = {astro-ph.CO},
       adsurl = {https://ui.adsabs.harvard.edu/abs/2026arXiv260320163Q}
}

@ARTICLE{iacconi25,
       author = {{Iacconi}, Laura and {Bhattacharya}, Sukannya and {Fasiello}, Matteo and {Wands}, David},
        title = "{Closing in on $α$-attractors}",
      journal = {arXiv e-prints},
         year = 2025,
        month = nov,
          eid = {arXiv:2511.14673},
        pages = {arXiv:2511.14673},
          doi = {10.48550/arXiv.2511.14673},
archivePrefix = {arXiv},
       eprint = {2511.14673},
 primaryClass = {astro-ph.CO},
       adsurl = {https://ui.adsabs.harvard.edu/abs/2025arXiv251114673I}
}

@ARTICLE{so_bb_new,
       author = {{The Simons Observatory Collaboration} and {Abril-Cabezas}, I. and {Adachi}, S. and {Ade}, P. and {Adler}, A.~E. and {Agrawal}, P. and {Aguirre}, J. and {Aiola}, S. and {Alford}, T. and {Ali}, A. and {Alonso}, D. and {Alvarez}, M.~A. and {An}, R. and {Aravena}, M. and {Arnold}, K. and {Ashton}, P. and {Astori}, F. and {Atkins}, Z. and {Austermann}, J. and {Azzoni}, S. and {Baccigalupi}, C. and {Baker}, D. and {Balafendiev}, R. and {Baleato Lizancos}, A. and {Barron}, D. and {Barry}, P. and {Bartlett}, J. and {Basyrov}, A. and {Battaglia}, N. and {Battistelli}, E.~S. and {Battye}, R. and {Bayer}, A. and {Bazarko}, A. and {Beall}, J.~A. and {Bean}, R. and {Beck}, D. and {Beckman}, S. and {Begin}, J. and {Beheshti}, A. and {Beringue}, B. and {Bhandarkar}, T. and {Bhimani}, S. and {Bianchini}, F. and {Biermann}, E. and {Billi}, M. and {Biquard}, S. and {Bixler}, B. and {Bizzarri}, L. and {Boada}, S. and {Boettger}, D. and {Bolliet}, B. and {Bond}, J.~R. and {Borrill}, J. and {Borrow}, J. and {Braithwaite}, C. and {Brien}, T.~L.~R. and {Brown}, M.~L. and {Bruno}, S.~M. and {Bryan}, S. and {Bustos}, R. and {Cai}, H. and {Calabrese}, E. and {Calafut}, V. and {Carl}, F.~M. and {Carones}, A. and {Carron}, J. and {Challinor}, A. and {Chamberlain}, E. and {Chanial}, P. and {Chen}, N. and {Cheung}, K. and {Chiang}, B. and {Chinone}, Y. and {Chluba}, J. and {Cho}, H.~S. and {Choi}, S.~K. and {Chu}, M. and {Clancy}, J. and {Clark}, S.~E. and {Clarke}, P. and {Cleary}, J. and {Clements}, D.~L. and {Connors}, J. and {Contaldi}, C. and {Coppi}, G. and {Corbett}, L. and {Cothard}, N.~F. and {Coulton}, W. and {Crichton}, D. and {Crowley}, K.~D. and {Crowley}, K.~T. and {Cukierman}, A. and {D'Ewart}, J.~M. and {Dachlythra}, K. and {Darwish}, O. and {Datta}, R. and {Day-Weiss}, S. and {de Haan}, T. and {Desai}, S. and {Devlin}, M. and {Di Mascolo}, L. and {Dicker}, S. and {Ding}, K. and {Doux}, C. and {Dow}, P. and {Doyle}, S. and {Duell}, C.~J. and {Duff}, S.~M. and {Duivenvoorden}, A.~J. and {Dunkley}, J. and {Duparc}, M. and {Dutcher}, D. and {D{\"u}nner}, R. and {Edenton}, M. and {El Bouhargani}, H. and {Embil Villagra}, C. and {Errard}, J. and {Fabbian}, G. and {Fanfani}, V. and {Farhadi Khouzani}, F. and {Farren}, G.~S. and {Fergusson}, J. and {Ferraro}, S. and {Flauger}, R. and {Forconi}, M. and {Foster}, A. and {Freese}, K. and {Frisch}, J.~C. and {Frolov}, A. and {Fuller}, G. and {Galitzki}, N. and {Gallardo}, P.~A. and {Galloni}, G. and {Galvez Ghersi}, J.~T. and {Ganga}, K. and {Garrido}, X. and {Gawiser}, E. and {Gerbino}, M. and {Gerras}, R. and {Giardiello}, S. and {Gill}, A. and {Gilles}, V. and {Giri}, U. and {Gleave}, E. and {Gluscevic}, V. and {Goeckner-Wald}, N. and {Goldstein}, S. and {Golec}, J.~E. and {Gordon}, S. and {Gralla}, M. and {Gratton}, S. and {Green}, D. and {Groh}, J.~C. and {Groppi}, C. and {Grubb}, S. and {Guan}, Y. and {Gupta}, N. and {Gu{\dh}mundsson}, J.~E. and {Hadzhiyska}, B. and {Hagstotz}, S. and {Hargrave}, P. and {Haridas}, S. and {Harrington}, K. and {Harrison}, I. and {Hasegawa}, M. and {Hasselfield}, M. and {Haynes}, V. and {Hazumi}, M. and {He}, A. and {Healy}, E. and {Henderson}, S.~W. and {Hensley}, B.~S. and {Hertig}, E. and {Herv{\'\i}as-Caimapo}, C. and {Higuchi}, M. and {Hill}, C.~A. and {Hill}, J.~C. and {Hilton}, M. and {Hincks}, A.~D. and {Hinshaw}, G. and {Hlo{\v{z}}ek}, R. and {Ho}, A.~Y.~Q. and {Ho}, S. and {Ho}, S.~P. and {Hoang}, T.~D. and {Hoh}, J. and {Holder}, J. and {Hood}, J. and {Hornecker}, E. and {Hornsby}, A.~L. and {Hotinli}, S.~C. and {Huang}, Z. and {Huber}, Z.~B. and {Hubmayr}, J. and {Huffenberger}, K. and {Hughes}, A. and {Hughes}, J.~P. and {Idicherian Lonappan}, A. and {Ikape}, M. and {Inaba}, K.},
        title = "{The Simons Observatory: forecasted constraints on primordial gravitational waves with the expanded array of Small Aperture Telescopes}",
      journal = {arXiv e-prints},
         year = 2025,
        month = dec,
          eid = {arXiv:2512.15833},
        pages = {arXiv:2512.15833},
          doi = {10.48550/arXiv.2512.15833},
archivePrefix = {arXiv},
       eprint = {2512.15833},
 primaryClass = {astro-ph.CO},
       adsurl = {https://ui.adsabs.harvard.edu/abs/2025arXiv251215833T}
}

@ARTICLE{kallosh25s,
       author = {{Kallosh}, Renata and {Linde}, Andrei},
        title = "{Singular $α$-attractors}",
      journal = {arXiv e-prints},
         year = 2025,
        month = dec,
          eid = {arXiv:2512.02969},
        pages = {arXiv:2512.02969},
          doi = {10.48550/arXiv.2512.02969},
archivePrefix = {arXiv},
       eprint = {2512.02969},
 primaryClass = {hep-th},
       adsurl = {https://ui.adsabs.harvard.edu/abs/2025arXiv251202969K}
}

@ARTICLE{paoletti22,
       author = {{Paoletti}, Daniela and {Finelli}, Fabio and {Valiviita}, Jussi and {Hazumi}, Masashi},
        title = "{Planck and BICEP/Keck Array 2018 constraints on primordial gravitational waves and perspectives for future B -mode polarization measurements}",
      journal = {\prd},
         year = 2022,
        month = oct,
       volume = {106},
       number = {8},
          eid = {083528},
        pages = {083528},
          doi = {10.1103/PhysRevD.106.083528},
archivePrefix = {arXiv},
       eprint = {2208.10482},
 primaryClass = {astro-ph.CO},
       adsurl = {https://ui.adsabs.harvard.edu/abs/2022PhRvD.106h3528P}
}

@ARTICLE{kallosh25,
       author = {{Kallosh}, Renata and {Linde}, Andrei},
        title = "{On the present status of inflationary cosmology}",
      journal = {General Relativity and Gravitation},
         year = 2025,
        month = sep,
       volume = {57},
       number = {10},
          eid = {135},
        pages = {135},
          doi = {10.1007/s10714-025-03470-6},
archivePrefix = {arXiv},
       eprint = {2505.13646},
 primaryClass = {hep-th},
       adsurl = {https://ui.adsabs.harvard.edu/abs/2025GReGr..57..135K}
}

@ARTICLE{tristram25,
       author = {{Tristram}, M. and {Douspis}, M. and {Gorce}, A. and {Henrot-Versill{\'e}}, S. and {Hergt}, L.~T. and {Ilic}, S. and {McBride}, L. and {Mu{\~n}oz-Echeverr{\'\i}a}, M. and {Pointecouteau}, E. and {Salvati}, L.},
        title = "{Combining CMB datasets with consistent foreground modelling}",
      journal = {arXiv e-prints},
         year = 2025,
        month = nov,
          eid = {arXiv:2511.04733},
        pages = {arXiv:2511.04733},
          doi = {10.48550/arXiv.2511.04733},
archivePrefix = {arXiv},
       eprint = {2511.04733},
 primaryClass = {astro-ph.CO},
       adsurl = {https://ui.adsabs.harvard.edu/abs/2025arXiv251104733T}
}

@ARTICLE{beringue25,
       author = {{Beringue}, Benjamin and {Surrao}, Kristen M. and {Hill}, J. Colin and {Atkins}, Zachary and {Battaglia}, Nicholas and {Bolliet}, Boris and {Calabrese}, Erminia and {Choi}, Steve K. and {Clark}, Susan E. and {Duivenvoorden}, Adriaan J. and {Dunkley}, Jo and {Giardiello}, Serena and {Goldstein}, Samuel and {Hensley}, Brandon S. and {Hlo{\v{z}}ek}, Ren{\'e}e and {Jense}, Hidde T. and {Kramer}, Darby and {La Posta}, Adrien and {Louis}, Thibaut and {Mehta}, Yogesh and {Moodley}, Kavilan and {Naess}, Sigurd and {Partridge}, Bruce and {Qu}, Frank J. and {Ried Guachalla}, Bernardita and {Sehgal}, Neelima and {Sif{\'o}n}, Crist{\'o}bal and {Staggs}, Suzanne T. and {Trac}, Hy and {Van Engelen}, Alexander and {Wollack}, Edward J.},
        title = "{The Atacama Cosmology Telescope: DR6 power spectrum foreground model and validation}",
      journal = {\jcap},
         year = 2025,
        month = oct,
       volume = {2025},
       number = {10},
          eid = {082},
        pages = {082},
          doi = {10.1088/1475-7516/2025/10/082},
archivePrefix = {arXiv},
       eprint = {2506.06274},
 primaryClass = {astro-ph.CO},
       adsurl = {https://ui.adsabs.harvard.edu/abs/2025JCAP...10..082B}
}

@ARTICLE{planck18-10,
       author = {{Planck Collaboration} and {Akrami}, Y. and {Arroja}, F. and {Ashdown}, M. and {Aumont}, J. and {Baccigalupi}, C. and {Ballardini}, M. and {Banday}, A.~J. and {Barreiro}, R.~B. and {Bartolo}, N. and {Basak}, S. and {Benabed}, K. and {Bernard}, J.-P. and {Bersanelli}, M. and {Bielewicz}, P. and {Bock}, J.~J. and {Bond}, J.~R. and {Borrill}, J. and {Bouchet}, F.~R. and {Boulanger}, F. and {Bucher}, M. and {Burigana}, C. and {Butler}, R.~C. and {Calabrese}, E. and {Cardoso}, J.-F. and {Carron}, J. and {Challinor}, A. and {Chiang}, H.~C. and {Colombo}, L.~P.~L. and {Combet}, C. and {Contreras}, D. and {Crill}, B.~P. and {Cuttaia}, F. and {de Bernardis}, P. and {de Zotti}, G. and {Delabrouille}, J. and {Delouis}, J.-M. and {Di Valentino}, E. and {Diego}, J.~M. and {Donzelli}, S. and {Dor{\'e}}, O. and {Douspis}, M. and {Ducout}, A. and {Dupac}, X. and {Dusini}, S. and {Efstathiou}, G. and {Elsner}, F. and {En{\ss}lin}, T.~A. and {Eriksen}, H.~K. and {Fantaye}, Y. and {Fergusson}, J. and {Fernandez-Cobos}, R. and {Finelli}, F. and {Forastieri}, F. and {Frailis}, M. and {Franceschi}, E. and {Frolov}, A. and {Galeotta}, S. and {Galli}, S. and {Ganga}, K. and {Gauthier}, C. and {G{\'e}nova-Santos}, R.~T. and {Gerbino}, M. and {Ghosh}, T. and {Gonz{\'a}lez-Nuevo}, J. and {G{\'o}rski}, K.~M. and {Gratton}, S. and {Gruppuso}, A. and {Gudmundsson}, J.~E. and {Hamann}, J. and {Handley}, W. and {Hansen}, F.~K. and {Herranz}, D. and {Hivon}, E. and {Hooper}, D.~C. and {Huang}, Z. and {Jaffe}, A.~H. and {Jones}, W.~C. and {Keih{\"a}nen}, E. and {Keskitalo}, R. and {Kiiveri}, K. and {Kim}, J. and {Kisner}, T.~S. and {Krachmalnicoff}, N. and {Kunz}, M. and {Kurki-Suonio}, H. and {Lagache}, G. and {Lamarre}, J.-M. and {Lasenby}, A. and {Lattanzi}, M. and {Lawrence}, C.~R. and {Le Jeune}, M. and {Lesgourgues}, J. and {Levrier}, F. and {Lewis}, A. and {Liguori}, M. and {Lilje}, P.~B. and {Lindholm}, V. and {L{\'o}pez-Caniego}, M. and {Lubin}, P.~M. and {Ma}, Y.-Z. and {Mac{\'\i}as-P{\'e}rez}, J.~F. and {Maggio}, G. and {Maino}, D. and {Mandolesi}, N. and {Mangilli}, A. and {Marcos-Caballero}, A. and {Maris}, M. and {Martin}, P.~G. and {Mart{\'\i}nez-Gonz{\'a}lez}, E. and {Matarrese}, S. and {Mauri}, N. and {McEwen}, J.~D. and {Meerburg}, P.~D. and {Meinhold}, P.~R. and {Melchiorri}, A. and {Mennella}, A. and {Migliaccio}, M. and {Mitra}, S. and {Miville-Desch{\^e}nes}, M.-A. and {Molinari}, D. and {Moneti}, A. and {Montier}, L. and {Morgante}, G. and {Moss}, A. and {M{\"u}nchmeyer}, M. and {Natoli}, P. and {N{\o}rgaard-Nielsen}, H.~U. and {Pagano}, L. and {Paoletti}, D. and {Partridge}, B. and {Patanchon}, G. and {Peiris}, H.~V. and {Perrotta}, F. and {Pettorino}, V. and {Piacentini}, F. and {Polastri}, L. and {Polenta}, G. and {Puget}, J.-L. and {Rachen}, J.~P. and {Reinecke}, M. and {Remazeilles}, M. and {Renzi}, A. and {Rocha}, G. and {Rosset}, C. and {Roudier}, G. and {Rubi{\~n}o-Mart{\'\i}n}, J.~A. and {Ruiz-Granados}, B. and {Salvati}, L. and {Sandri}, M. and {Savelainen}, M. and {Scott}, D. and {Shellard}, E.~P.~S. and {Shiraishi}, M. and {Sirignano}, C. and {Sirri}, G. and {Spencer}, L.~D. and {Sunyaev}, R. and {Suur-Uski}, A.-S. and {Tauber}, J.~A. and {Tavagnacco}, D. and {Tenti}, M. and {Toffolatti}, L. and {Tomasi}, M. and {Trombetti}, T. and {Valiviita}, J. and {Van Tent}, B. and {Vielva}, P. and {Villa}, F. and {Vittorio}, N. and {Wandelt}, B.~D. and {Wehus}, I.~K. and {White}, S.~D.~M. and {Zacchei}, A. and {Zibin}, J.~P. and {Zonca}, A.},
        title = "{Planck 2018 results. X. Constraints on inflation}",
      journal = {\aap},
         year = 2020,
        month = sep,
       volume = {641},
          eid = {A10},
        pages = {A10},
          doi = {10.1051/0004-6361/201833887},
archivePrefix = {arXiv},
       eprint = {1807.06211},
 primaryClass = {astro-ph.CO},
       adsurl = {https://ui.adsabs.harvard.edu/abs/2020A&A...641A..10P}
}

@ARTICLE{inflpostplanck13,
       author = {{Linde}, Andrei},
        title = "{Inflationary Cosmology after Planck 2013}",
      journal = {arXiv e-prints},
         year = 2014,
        month = feb,
          eid = {arXiv:1402.0526},
        pages = {arXiv:1402.0526},
          doi = {10.48550/arXiv.1402.0526},
archivePrefix = {arXiv},
       eprint = {1402.0526},
 primaryClass = {hep-th},
       adsurl = {https://ui.adsabs.harvard.edu/abs/2014arXiv1402.0526L}
}

@ARTICLE{jense25planck,
       author = {{Jense}, Hidde and {Vi{\~n}a}, Marc and {Calabrese}, Erminia and {Hill}, Colin},
        title = "{The choice of Planck CMB likelihood in cosmological analyses}",
      journal = {arXiv e-prints},
         year = 2025,
        month = oct,
          eid = {arXiv:2510.09430},
        pages = {arXiv:2510.09430},
          doi = {10.48550/arXiv.2510.09430},
archivePrefix = {arXiv},
       eprint = {2510.09430},
 primaryClass = {astro-ph.CO},
       adsurl = {https://ui.adsabs.harvard.edu/abs/2025arXiv251009430J}
}

@ARTICLE{ellis25,
       author = {{Ellis}, John and {Garcia}, Marcos A.~G. and {Olive}, Keith A. and {Verner}, Sarunas},
        title = "{Constraints on Attractor Models of Inflation and Reheating from Planck, BICEP/Keck, ACT DR6, and SPT-3G Data}",
      journal = {arXiv e-prints},
         year = 2025,
        month = oct,
          eid = {arXiv:2510.18656},
        pages = {arXiv:2510.18656},
          doi = {10.48550/arXiv.2510.18656},
archivePrefix = {arXiv},
       eprint = {2510.18656},
 primaryClass = {hep-ph},
       adsurl = {https://ui.adsabs.harvard.edu/abs/2025arXiv251018656E}
}

@ARTICLE{sroll2,
       author = {{Pagano}, L. and {Delouis}, J.-M. and {Mottet}, S. and {Puget}, J.-L. and {Vibert}, L.},
        title = "{Reionization optical depth determination from Planck HFI data with ten percent accuracy}",
      journal = {\aap},
         year = 2020,
        month = mar,
       volume = {635},
          eid = {A99},
        pages = {A99},
          doi = {10.1051/0004-6361/201936630},
archivePrefix = {arXiv},
       eprint = {1908.09856},
 primaryClass = {astro-ph.CO},
       adsurl = {https://ui.adsabs.harvard.edu/abs/2020A&A...635A..99P}
}

@book{weinbergcosmo,
    author = "Weinberg, Steven",
    title = "{Cosmology}",
    isbn = "978-0-19-852682-7",
    year = "2008"
}

@book{baumanncosmo,
    author = "Baumann, Daniel",
    title = "{Cosmology}",
    doi = "10.1017/9781108937092",
    isbn = "978-1-108-93709-2, 978-1-108-83807-8",
    publisher = "Cambridge University Press",
    month = "7",
    year = "2022"
}

@book{liddlecosmo,
    author = "Liddle, Andrew R.",
    title = "{An introduction to modern cosmology}",
    year = "1998"
}

@book{moderncosmo,
    author = "Dodelson, Scott",
    title = "{Modern Cosmology}",
    isbn = "978-0-12-219141-1",
    publisher = "Academic Press",
    address = "Amsterdam",
    year = "2003"
}

@ARTICLE{getdist25,
       author = {{Lewis}, Antony},
        title = "{GetDist: a Python package for analysing Monte Carlo samples}",
      journal = {\jcap},
         year = 2025,
        month = aug,
       volume = {2025},
       number = {8},
          eid = {025},
        pages = {025},
          doi = {10.1088/1475-7516/2025/08/025},
archivePrefix = {arXiv},
       eprint = {1910.13970},
 primaryClass = {astro-ph.IM},
       adsurl = {https://ui.adsabs.harvard.edu/abs/2025JCAP...08..025L}
}

@ARTICLE{cartis18b,
       author = {{Cartis}, Coralia and {Roberts}, Lindon and {Sheridan-Methven}, Oliver},
        title = "{Escaping local minima with derivative-free methods: a numerical investigation}",
      journal = {arXiv e-prints},
         year = 2018,
        month = dec,
          eid = {arXiv:1812.11343},
        pages = {arXiv:1812.11343},
          doi = {10.48550/arXiv.1812.11343},
archivePrefix = {arXiv},
       eprint = {1812.11343},
 primaryClass = {math.OC},
       adsurl = {https://ui.adsabs.harvard.edu/abs/2018arXiv181211343C}
}

@ARTICLE{cartis18,
       author = {{Cartis}, Coralia and {Fiala}, Jan and {Marteau}, Benjamin and {Roberts}, Lindon},
        title = "{Improving the Flexibility and Robustness of Model-Based Derivative-Free Optimization Solvers}",
      journal = {arXiv e-prints},
         year = 2018,
        month = mar,
          eid = {arXiv:1804.00154},
        pages = {arXiv:1804.00154},
          doi = {10.48550/arXiv.1804.00154},
archivePrefix = {arXiv},
       eprint = {1804.00154},
 primaryClass = {math.OC},
       adsurl = {https://ui.adsabs.harvard.edu/abs/2018arXiv180400154C}
}

@ARTICLE{ferreira25,
       author = {{Ferreira}, Elisa G.~M. and {McDonough}, Evan and {Balkenhol}, Lennart and {Kallosh}, Renata and {Knox}, Lloyd and {Linde}, Andrei},
        title = "{The BAO-CMB Tension and Implications for Inflation}",
      journal = {arXiv e-prints},
         year = 2025,
        month = jul,
          eid = {arXiv:2507.12459},
        pages = {arXiv:2507.12459},
          doi = {10.48550/arXiv.2507.12459},
archivePrefix = {arXiv},
       eprint = {2507.12459},
 primaryClass = {astro-ph.CO},
       adsurl = {https://ui.adsabs.harvard.edu/abs/2025arXiv250712459F}
}

@ARTICLE{sofc2025,
       author = {{Abitbol}, M. and {Abril-Cabezas}, I. and {Adachi}, S. and {Ade}, P. and {Adler}, A.~E. and {Agrawal}, P. and {Aguirre}, J. and {Ahmed}, Z. and {Aiola}, S. and {Alford}, T. and {Ali}, A. and {Alonso}, D. and {Alvarez}, M.~A. and {An}, R. and {Arnold}, K. and {Ashton}, P. and {Atkins}, Z. and {Austermann}, J. and {Azzoni}, S. and {Baccigalupi}, C. and {Baleato Lizancos}, A. and {Barron}, D. and {Barry}, P. and {Bartlett}, J. and {Battaglia}, N. and {Battye}, R. and {Baxter}, E. and {Bazarko}, A. and {Beall}, J.~A. and {Bean}, R. and {Beck}, D. and {Beckman}, S. and {Begin}, J. and {Beheshti}, A. and {Beringue}, B. and {Bhandarkar}, T. and {Bhimani}, S. and {Bianchini}, F. and {Biermann}, E. and {Biquard}, S. and {Bixler}, B. and {Boada}, S. and {Boettger}, D. and {Bolliet}, B. and {Bond}, J.~R. and {Borrill}, J. and {Borrow}, J. and {Braithwaite}, C. and {Brien}, T.~L.~R. and {Brown}, M.~L. and {Bruno}, S.~M. and {Bryan}, S. and {Bustos}, R. and {Cai}, H. and {Calabrese}, E. and {Calafut}, V. and {Carl}, F.~M. and {Carones}, A. and {Carron}, J. and {Challinor}, A. and {Chanial}, P. and {Chen}, N. and {Cheung}, K. and {Chiang}, B. and {Chinone}, Y. and {Chluba}, J. and {Cho}, H.~S. and {Choi}, S.~K. and {Chu}, M. and {Clancy}, J. and {Clark}, S.~E. and {Clarke}, P. and {Cleary}, J. and {Clements}, D.~L. and {Connors}, J. and {Contaldi}, C. and {Coppi}, G. and {Corbett}, L. and {Cothard}, N.~F. and {Coulton}, W. and {Crowley}, K.~D. and {Crowley}, K.~T. and {Cukierman}, A. and {D'Ewart}, J.~M. and {Dachlythra}, K. and {Datta}, R. and {Day-Weiss}, S. and {de Haan}, T. and {Devlin}, M. and {Di Mascolo}, L. and {Dicker}, S. and {Dober}, B. and {Doux}, C. and {Dow}, P. and {Doyle}, S. and {Duell}, C.~J. and {Duff}, S.~M. and {Duivenvoorden}, A.~J. and {Dunkley}, J. and {Dutcher}, D. and {D{\"u}nner}, R. and {Edenton}, M. and {El Bouhargani}, H. and {Errard}, J. and {Fabbian}, G. and {Fanfani}, V. and {Farren}, G.~S. and {Fergusson}, J. and {Ferraro}, S. and {Flauger}, R. and {Foster}, A. and {Freese}, K. and {Frisch}, J.~C. and {Frolov}, A. and {Fuller}, G. and {Galitzki}, N. and {Gallardo}, P.~A. and {Galvez Ghersi}, J.~T. and {Ganga}, K. and {Gao}, J. and {Garrido}, X. and {Gawiser}, E. and {Gerbino}, M. and {Gerras}, R. and {Giardiello}, S. and {Gill}, A. and {Gilles}, V. and {Giri}, U. and {Gleave}, E. and {Gluscevic}, V. and {Goeckner-Wald}, N. and {Golec}, J.~E. and {Gordon}, S. and {Gralla}, M. and {Gratton}, S. and {Green}, D. and {Groh}, J.~C. and {Groppi}, C. and {Guan}, Y. and {Gupta}, N. and {Gudmundsson}, J.~E. and {Hagstotz}, S. and {Hargrave}, P. and {Haridas}, S. and {Harrington}, K. and {Harrison}, I. and {Hasegawa}, M. and {Hasselfield}, M. and {Haynes}, V. and {Hazumi}, M. and {He}, A. and {Healy}, E. and {Henderson}, S.~W. and {Hensley}, B.~S. and {Hertig}, E. and {Herv{\'\i}as-Caimapo}, C. and {Higuchi}, M. and {Hill}, C.~A. and {Hill}, J.~C. and {Hilton}, G. and {Hilton}, M. and {Hincks}, A.~D. and {Hinshaw}, G. and {Hlo{\v{z}}ek}, R. and {Ho}, A.~Y.~Q. and {Ho}, S. and {Ho}, S.~P. and {Hoang}, T.~D. and {Hoh}, J. and {Hornecker}, E. and {Hornsby}, A.~L. and {Hotinli}, S.~C. and {Huang}, Z. and {Huber}, Z.~B. and {Hubmayr}, J. and {Huffenberger}, K. and {Hughes}, J.~P. and {Idicherian Lonappan}, A. and {Ikape}, M. and {Irwin}, K. and {Iuliano}, J. and {Jaffe}, A.~H. and {Jain}, B. and {Jense}, H.~T. and {Jeong}, O. and {Johnson}, A. and {Johnson}, B.~R. and {Johnson}, M. and {Jones}, M. and {Jost}, B. and {Kaneko}, D. and {Karpel}, E.~D. and {Kasai}, Y. and {Katayama}, N. and {Keating}, B. and {Keller}, B. and {Keskitalo}, R. and {Kim}, J. and {Kisner}, T. and {Kiuchi}, K.},
        title = "{The Simons Observatory: science goals and forecasts for the enhanced Large Aperture Telescope}",
      journal = {\jcap},
         year = 2025,
        month = aug,
       volume = {2025},
       number = {8},
          eid = {034},
        pages = {034},
          doi = {10.1088/1475-7516/2025/08/034},
archivePrefix = {arXiv},
       eprint = {2503.00636},
 primaryClass = {astro-ph.IM},
       adsurl = {https://ui.adsabs.harvard.edu/abs/2025JCAP...08..034A}
}

@ARTICLE{vitrier25,
       author = {{Vitrier}, A. and {Fichman}, K. and {Balkenhol}, L. and {Camphuis}, E. and {Guidi}, F. and {Khalife}, A.~R. and {Anderson}, A.~J. and {Ansarinejad}, B. and {Archipley}, M. and {Benabed}, K. and {Bender}, A.~N. and {Benson}, B.~A. and {Bianchini}, F. and {Bleem}, L.~E. and {Bouchet}, F.~R. and {Bryant}, L. and {Campitiello}, M.~G. and {Carlstrom}, J.~E. and {Chang}, C.~L. and {Chaubal}, P. and {Chichura}, P.~M. and {Chokshi}, A. and {Chou}, T.-L. and {Coerver}, A. and {Crawford}, T.~M. and {Daley}, C. and {de Haan}, T. and {Dibert}, K.~R. and {Dobbs}, M.~A. and {Doohan}, M. and {Doussot}, A. and {Dutcher}, D. and {Everett}, W. and {Feng}, C. and {Ferguson}, K.~R. and {Ferree}, N.~C. and {Foster}, A. and {Galli}, S. and {Gambrel}, A.~E. and {Gardner}, R.~W. and {Ge}, F. and {Goeckner-Wald}, N. and {Gualtieri}, R. and {Guns}, S. and {Halverson}, N.~W. and {Hivon}, E. and {Holder}, G.~P. and {Holzapfel}, W.~L. and {Hood}, J.~C. and {Hryciuk}, A. and {Huang}, N. and {K{\'e}ruzor{\'e}}, F. and {Knox}, L. and {Korman}, M. and {Kornoelje}, K. and {Kuo}, C.-L. and {Levy}, K. and {Li}, Y. and {Lowitz}, A.~E. and {Lu}, C. and {Lynch}, G.~P. and {Maniyar}, A. and {Martsen}, E.~S. and {Menanteau}, F. and {Millea}, M. and {Montgomery}, J. and {Nakato}, Y. and {Natoli}, T. and {Noble}, G.~I. and {Omori}, Y. and {Ouellette}, A. and {Pan}, Z. and {Paschos}, P. and {Phadke}, K.~A. and {Pollak}, A.~W. and {Prabhu}, K. and {Quan}, W. and {Rahimi}, M. and {Rahlin}, A. and {Reichardt}, C.~L. and {Rouble}, M. and {Ruhl}, J.~E. and {Schiappucci}, E. and {Silva Oliveira}, A.~C. and {Simpson}, A. and {Sobrin}, J.~A. and {Stark}, A.~A. and {Stephen}, J. and {Tandoi}, C. and {Thorne}, B. and {Trendafilova}, C. and {Umilta}, C. and {Vieira}, J.~D. and {Wan}, Y. and {Whitehorn}, N. and {Wu}, W.~L.~K. and {Young}, M.~R. and {Zebrowski}, J.~A.},
        title = "{Towards constraining cosmological parameters with SPT-3G observations of 25\% of the sky}",
      journal = {arXiv e-prints},
         year = 2025,
        month = oct,
          eid = {arXiv:2510.24669},
        pages = {arXiv:2510.24669},
          doi = {10.48550/arXiv.2510.24669},
archivePrefix = {arXiv},
       eprint = {2510.24669},
 primaryClass = {astro-ph.CO},
       adsurl = {https://ui.adsabs.harvard.edu/abs/2025arXiv251024669V}
}

@ARTICLE{lb23,
       author = {{LiteBIRD Collaboration} and {Allys}, E. and {Arnold}, K. and {Aumont}, J. and {Aurlien}, R. and {Azzoni}, S. and {Baccigalupi}, C. and {Banday}, A.~J. and {Banerji}, R. and {Barreiro}, R.~B. and {Bartolo}, N. and {Bautista}, L. and {Beck}, D. and {Beckman}, S. and {Bersanelli}, M. and {Boulanger}, F. and {Brilenkov}, M. and {Bucher}, M. and {Calabrese}, E. and {Campeti}, P. and {Carones}, A. and {Casas}, F.~J. and {Catalano}, A. and {Chan}, V. and {Cheung}, K. and {Chinone}, Y. and {Clark}, S.~E. and {Columbro}, F. and {D'Alessandro}, G. and {de Bernardis}, P. and {de Haan}, T. and {de la Hoz}, E. and {De Petris}, M. and {Torre}, S. Della and {Diego-Palazuelos}, P. and {Dobbs}, M. and {Dotani}, T. and {Duval}, J.~M. and {Elleflot}, T. and {Eriksen}, H.~K. and {Errard}, J. and {Essinger-Hileman}, T. and {Finelli}, F. and {Flauger}, R. and {Franceschet}, C. and {Fuskeland}, U. and {Galloway}, M. and {Ganga}, K. and {Gerbino}, M. and {Gervasi}, M. and {G{\'e}nova-Santos}, R.~T. and {Ghigna}, T. and {Giardiello}, S. and {Gjerl{\o}w}, E. and {Grain}, J. and {Grupp}, F. and {Gruppuso}, A. and {Gudmundsson}, J.~E. and {Halverson}, N.~W. and {Hargrave}, P. and {Hasebe}, T. and {Hasegawa}, M. and {Hazumi}, M. and {Henrot-Versill{\'e}}, S. and {Hensley}, B. and {Hergt}, L.~T. and {Herman}, D. and {Hivon}, E. and {Hlozek}, R.~A. and {Hornsby}, A.~L. and {Hoshino}, Y. and {Hubmayr}, J. and {Ichiki}, K. and {Iida}, T. and {Imada}, H. and {Ishino}, H. and {Jaehnig}, G. and {Katayama}, N. and {Kato}, A. and {Keskitalo}, R. and {Kisner}, T. and {Kobayashi}, Y. and {Kogut}, A. and {Kohri}, K. and {Komatsu}, E. and {Komatsu}, K. and {Konishi}, K. and {Krachmalnicoff}, N. and {Kuo}, C.~L. and {Lamagna}, L. and {Lattanzi}, M. and {Lee}, A.~T. and {Leloup}, C. and {Levrier}, F. and {Linder}, E. and {Luzzi}, G. and {Macias-Perez}, J. and {Maciaszek}, T. and {Maffei}, B. and {Maino}, D. and {Mandelli}, S. and {Mart{\'\i}nez-Gonz{\'a}lez}, E. and {Masi}, S. and {Massa}, M. and {Matarrese}, S. and {Matsuda}, F.~T. and {Matsumura}, T. and {Mele}, L. and {Migliaccio}, M. and {Minami}, Y. and {Moggi}, A. and {Montgomery}, J. and {Montier}, L. and {Morgante}, G. and {Mot}, B. and {Nagano}, Y. and {Nagasaki}, T. and {Nagata}, R. and {Nakano}, R. and {Namikawa}, T. and {Nati}, F. and {Natoli}, P. and {Nerval}, S. and {Noviello}, F. and {Odagiri}, K. and {Oguri}, S. and {Ohsaki}, H. and {Pagano}, L. and {Paiella}, A. and {Paoletti}, D. and {Passerini}, A. and {Patanchon}, G. and {Piacentini}, F. and {Piat}, M. and {Pisano}, G. and {Polenta}, G. and {Poletti}, D. and {Prouv{\'e}}, T. and {Puglisi}, G. and {Rambaud}, D. and {Raum}, C. and {Realini}, S. and {Reinecke}, M. and {Remazeilles}, M. and {Ritacco}, A. and {Roudil}, G. and {Rubino-Martin}, J.~A. and {Russell}, M. and {Sakurai}, H. and {Sakurai}, Y. and {Sasaki}, M. and {Scott}, D. and {Sekimoto}, Y. and {Shinozaki}, K. and {Shiraishi}, M. and {Shirron}, P. and {Signorelli}, G. and {Spinella}, F. and {Stever}, S. and {Stompor}, R. and {Sugiyama}, S. and {Sullivan}, R.~M. and {Suzuki}, A. and {Svalheim}, T.~L. and {Switzer}, E. and {Takaku}, R. and {Takakura}, H. and {Takase}, Y. and {Tartari}, A. and {Terao}, Y. and {Thermeau}, J. and {Thommesen}, H. and {Thompson}, K.~L. and {Tomasi}, M. and {Tominaga}, M. and {Tristram}, M. and {Tsuji}, M. and {Tsujimoto}, M. and {Vacher}, L. and {Vielva}, P. and {Vittorio}, N. and {Wang}, W. and {Watanuki}, K. and {Wehus}, I.~K. and {Weller}, J. and {Westbrook}, B. and {Wilms}, J. and {Winter}, B. and {Wollack}, E.~J. and {Yumoto}, J. and {Zannoni}, M. and {Collaboration LiteB I R D}},
        title = "{Probing cosmic inflation with the LiteBIRD cosmic microwave background polarization survey}",
      journal = {Progress of Theoretical and Experimental Physics},
         year = 2023,
        month = apr,
       volume = {2023},
       number = {4},
          eid = {042F01},
        pages = {042F01},
          doi = {10.1093/ptep/ptac150},
archivePrefix = {arXiv},
       eprint = {2202.02773},
 primaryClass = {astro-ph.IM},
       adsurl = {https://ui.adsabs.harvard.edu/abs/2023PTEP.2023d2F01L}
}

@ARTICLE{camphuis25,
       author = {{Camphuis}, E. and {Quan}, W. and {Balkenhol}, L. and {Khalife}, A.~R. and {Ge}, F. and {Guidi}, F. and {Huang}, N. and {Lynch}, G.~P. and {Omori}, Y. and {Trendafilova}, C. and {Anderson}, A.~J. and {Ansarinejad}, B. and {Archipley}, M. and {Barry}, P.~S. and {Benabed}, K. and {Bender}, A.~N. and {Benson}, B.~A. and {Bianchini}, F. and {Bleem}, L.~E. and {Bouchet}, F.~R. and {Bryant}, L. and {Campitiello}, M.~G. and {Carlstrom}, J.~E. and {Chang}, C.~L. and {Chaubal}, P. and {Chichura}, P.~M. and {Chokshi}, A. and {Chou}, T. -L. and {Coerver}, A. and {Crawford}, T.~M. and {Daley}, C. and {de Haan}, T. and {Dibert}, K.~R. and {Dobbs}, M.~A. and {Doohan}, M. and {Doussot}, A. and {Dutcher}, D. and {Everett}, W. and {Feng}, C. and {Ferguson}, K.~R. and {Fichman}, K. and {Foster}, A. and {Galli}, S. and {Gambrel}, A.~E. and {Gardner}, R.~W. and {Goeckner-Wald}, N. and {Gualtieri}, R. and {Guns}, S. and {Halverson}, N.~W. and {Hivon}, E. and {Holder}, G.~P. and {Holzapfel}, W.~L. and {Hood}, J.~C. and {Hryciuk}, A. and {K{\'e}ruzor{\'e}}, F. and {Knox}, L. and {Korman}, M. and {Kornoelje}, K. and {Kuo}, C. -L. and {Levy}, K. and {Lowitz}, A.~E. and {Lu}, C. and {Maniyar}, A. and {Martsen}, E.~S. and {Menanteau}, F. and {Millea}, M. and {Montgomery}, J. and {Nakato}, Y. and {Natoli}, T. and {Noble}, G.~I. and {Ouellette}, A. and {Pan}, Z. and {Paschos}, P. and {Phadke}, K.~A. and {Pollak}, A.~W. and {Prabhu}, K. and {Raghunathan}, S. and {Rahimi}, M. and {Rahlin}, A. and {Reichardt}, C.~L. and {Rouble}, M. and {Ruhl}, J.~E. and {Schiappucci}, E. and {Simpson}, A. and {Sobrin}, J.~A. and {Stark}, A.~A. and {Stephen}, J. and {Tandoi}, C. and {Thorne}, B. and {Umilta}, C. and {Vieira}, J.~D. and {Vitrier}, A. and {Wan}, Y. and {Whitehorn}, N. and {Wu}, W.~L.~K. and {Young}, M.~R. and {Zebrowski}, J.~A.},
        title = "{SPT-3G D1: CMB temperature and polarization power spectra and cosmology from 2019 and 2020 observations of the SPT-3G Main field}",
      journal = {arXiv e-prints},
         year = 2025,
        month = jun,
          eid = {arXiv:2506.20707},
        pages = {arXiv:2506.20707},
          doi = {10.48550/arXiv.2506.20707},
archivePrefix = {arXiv},
       eprint = {2506.20707},
 primaryClass = {astro-ph.CO},
       adsurl = {https://ui.adsabs.harvard.edu/abs/2025arXiv250620707C}
}

@ARTICLE{abazajian16,
   author = {{Abazajian}, K.~N. and {Adshead}, P. and {Ahmed}, Z. and {Allen}, S.~W. and 
	{Alonso}, D. and {Arnold}, K.~S. and {Baccigalupi}, C. and {Bartlett}, J.~G. and 
	{Battaglia}, N. and {Benson}, B.~A. and {Bischoff}, C.~A. and 
	{Borrill}, J. and {Buza}, V. and {Calabrese}, E. and {Caldwell}, R. and 
	{Carlstrom}, J.~E. and {Chang}, C.~L. and {Crawford}, T.~M. and 
	{Cyr-Racine}, F.-Y. and {De Bernardis}, F. and {de Haan}, T. and 
	{di Serego Alighieri}, S. and {Dunkley}, J. and {Dvorkin}, C. and 
	{Errard}, J. and {Fabbian}, G. and {Feeney}, S. and {Ferraro}, S. and 
	{Filippini}, J.~P. and {Flauger}, R. and {Fuller}, G.~M. and 
	{Gluscevic}, V. and {Green}, D. and {Grin}, D. and {Grohs}, E. and 
	{Henning}, J.~W. and {Hill}, J.~C. and {Hlozek}, R. and {Holder}, G. and 
	{Holzapfel}, W. and {Hu}, W. and {Huffenberger}, K.~M. and {Keskitalo}, R. and 
	{Knox}, L. and {Kosowsky}, A. and {Kovac}, J. and {Kovetz}, E.~D. and 
	{Kuo}, C.-L. and {Kusaka}, A. and {Le Jeune}, M. and {Lee}, A.~T. and 
	{Lilley}, M. and {Loverde}, M. and {Madhavacheril}, M.~S. and 
	{Mantz}, A. and {Marsh}, D.~J.~E. and {McMahon}, J. and {Meerburg}, P.~D. and 
	{Meyers}, J. and {Miller}, A.~D. and {Munoz}, J.~B. and {Nguyen}, H.~N. and 
	{Niemack}, M.~D. and {Peloso}, M. and {Peloton}, J. and {Pogosian}, L. and 
	{Pryke}, C. and {Raveri}, M. and {Reichardt}, C.~L. and {Rocha}, G. and 
	{Rotti}, A. and {Schaan}, E. and {Schmittfull}, M.~M. and {Scott}, D. and 
	{Sehgal}, N. and {Shandera}, S. and {Sherwin}, B.~D. and {Smith}, T.~L. and 
	{Sorbo}, L. and {Starkman}, G.~D. and {Story}, K.~T. and {van Engelen}, A. and 
	{Vieira}, J.~D. and {Watson}, S. and {Whitehorn}, N. and {Kimmy Wu}, W.~L.
	},
    title = "{CMB-S4 Science Book, First Edition}",
  journal = {ArXiv e-prints},
archivePrefix = "arXiv",
   eprint = {1610.02743},
     year = 2016,
    month = oct,
   adsurl = {http://adsabs.harvard.edu/abs/2016arXiv161002743A}
}

@ARTICLE{lewis02b,
   author = {{Lewis}, A. and {Bridle}, S.},
    title = "{Cosmological parameters from CMB and other data: A Monte Carlo approach}",
  journal = {\prd},
     year = 2002,
    month = nov,
   volume = 66,
   number = 10,
    pages = {103511-+}
}

%%%%%%%%%%%%%%%%%%%%%%%%%%%%%%%%%%%%%%%%%%%%%%%%%%

\end{document}